\documentclass[a4paper,11pt]{article}
\usepackage{graphicx}
\usepackage{dcolumn}
\usepackage{bm}
\usepackage{amssymb}    
\usepackage{amsmath}
\usepackage{authblk}

\usepackage{bbm}
\usepackage[usenames]{color}
\usepackage{afterpage}
\usepackage{mathptmx}

\begin{document}

\title{Solitary states in complex networks: impact of topology}

\author[1]{Leonhard Sch\"ulen}
\author[2]{Maria Mikhailenko}
\author[3]{Everton S. Medeiros}
\author[1,4]{Anna Zakharova}
\affil[1]{ Institut f\"ur Theoretische Physik, Technische Universit\"at Berlin, Hardenbergstr. 36, 10623 Berlin, Germany}
\affil[2]{Charité – Universitätsmedizin Berlin, Einstein Center for Neurosciences Berlin, Charitéplatz 1, 10117, Berlin, Germany} 
\affil[3]{Institute for Chemistry and Biology of the Marine Environment, Carl-von-Ossietzky-University Oldenburg, Carl-von-Ossietzky-Straße 9 - 11, 26111 Oldenburg, Germany}
\affil[4]{Bernstein Center for Computational Neuroscience, Humboldt-Universit\"at zu Berlin, Philippstra\ss{}e 13, 10115 Berlin,
Germany}
\maketitle

\begin{abstract}
The dynamical behavior of networked systems is expected to reflect the features of their coupling structure. Yet, symmetry-broken solutions often occur in symmetrically coupled networks. An example is provided by the so-called solitary states where the dynamics of one network node is different from the entire symmetric network. Here, we investigate the structural constraints of networks for the appearance of solitary states in their dynamics. By performing a large number of numerical simulations, we find that such states occur with high probability in asymmetric networks, such as the ones exhibiting the scale-free property. Next, we analyze the structural features of the networks demonstrating solitary states to reveal that the minimum number of connections in the adjacent nodes of a solitary one is crucial for the appearance of the solitary states. Finally, we perform bifurcation analysis of a dimension-reduced system, confirming the influence of the connectivity of the neighboring nodes on the solitary ones.
\end{abstract}

\section{Introduction}\label{sec:intro}

Understanding the interplay between network topology and emerging dynamics is one of the central issues in the field of nonlinear dynamics and the theory of complex networks \cite{BOC06,PEC14,POE15}. Often, perfectly symmetric networked systems exhibit symmetry-broken solutions \cite{ABR2004,ZAK20}. For networks of oscillators, a particularly interesting example is the extreme case of cluster synchronization in which only one oscillator does not synchronize with the rest of the network. This highly unbalanced network configuration is called ``solitary'' state \cite{MAI14a,JAR15}. The occurrence of these states has been observed in a variety of symmetrically coupled systems such as globally coupled nonlinear oscillators \cite{MAI14a}, in non-locally (and globally) coupled networks of phase oscillators \cite{JAR18,HEL20}, coupled chaotic maps \cite{RYB17,RYB18,SEM18}, coupled excitable systems \cite{FRA22}, multilayer networks \cite{MIK19,SCH21,MAJ19,RYB21}, and time-delayed systems \cite{SCH19}. Such ubiquity naturally drew attention to solitary states and raised questions about the mechanism behind their appearance \cite{RYB19,SCH21}. Recently, Sch\"ulen et al. have addressed this issue for coupled neural oscillators and found that solitary states are created sub-critically in a fold-bifurcation \cite{SCH22}. Additionally, they have demonstrated that the dynamics of the solitary oscillator can also be chaotic while the synchronized cluster stays periodic \cite{SCH22}. 

Despite the advanced understanding of ubiquity and onset of the solitary states, some essential questions remain unanswered. For instance, what are the topological constraints for the emergence of these asymmetric solutions? Naturally, symmetry-broken solutions are unexpected in the scope of symmetric systems with uniform distribution of node degree. However, complex networks with asymmetries in their structure of connections, such as the ones with scale-free property presenting a power-law distribution of node degree, do not offer any immediate advantage for the occurrence of solitary states. In fact, it has been recently found that asymmetries can actually favor synchronization \cite{NIS16,HAR19} or regular behavior \cite{MED21,YAN22}. Therefore, the onset of solitary states in asymmetric networks is as equally important as the symmetric case, especially for a better understanding of how these states occur in general setups. 

In this work, we address the onset, and the topological dependencies of solitary states in complex networks of FitzHugh-Nagumo oscillators. We begin our investigation by studying the appearance of solitary states in three different network topologies: a symmetric nonlocally coupled ring, a random, and a scale-free graph. To establish the range of parameters in which the solitary states would appear, we use the symmetric nonlocally coupled ring as a baseline for the parameter search. We find that the random and the scale-free networks exhibit solitary states for similar parameter values as the symmetric one. These results are verified over an ensemble of realizations of such networks with different initial conditions. Further, we proceed to the topological analysis of the networks, aiming to identify a common feature of solitary nodes. We analyze their node degrees, neighbor node degrees, and eigenvector centrality. Among these measures, we found that a high neighbor node degree correlates with the onset of solitary states. Our next step is to verify this observation more rigorously by performing bifurcation analysis. For that, we consider geometrical arguments to obtain a dimension-reduced system, where the bifurcation analysis demonstrates the onset of the solitary states in a fold bifurcation and their dependencies on the minimum value of the neighbors node degree. 

\section{Solitary states in complex networks}\label{sec:system}

We consider different network topologies, each network consisting of identical FitzHugh-Nagumo (FHN) elements in the oscillatory regime. Such networks are numerically simulated using the following dimensionless equations:  
\begin{align}
    \varepsilon\frac{du_{i}}{dt}&=f(u_i, v_i) +\sum\limits_{j=1}^{N} \frac{A_{ij}}{d_{i}}[\sigma_u(u_{j}-u_{i})+ \sigma_{v}(v_{j}-v_{i})], \nonumber \\
\frac{dv_{i}}{dt}&=g(u_i, v_i),
\label{equ:fhn_network}
\end{align}
where the pair ($u_i$, $v_i$) accounts  for the activator and inhibitor variables, respectively, of each FHN oscillator $i$ with $i=1, \dots, N$. The parameter $N$ prescribes the network size. The functions:
\begin{equation}
 f(u_i, v_i)=u_i - \frac{u_i^3}{3} - v_i, \quad g(u_i, v_i)=g(u_i) = u_i +a
 \label{equ:functions}
\end{equation}
specify the local dynamics of the variables $u_i$ and $v_i$. The parameter $\varepsilon$ defines the time scale separation between the fast ($u_i$) and slow ($v_i$) variables. Throughout this study, we fix $\varepsilon=0.1$. For isolated oscillators, the threshold parameter $a$ separates the excitatory ($|a|>1$) from the oscillatory ($|a|<1$) regime through a supercritical Hopf bifurcation. We set each oscillator in the oscillatory regime by keeping $a=0.5$. The different network topologies are characterized by their adjacency matrix $A_{ij}$. The FHN oscillators are diffusively coupled via both variables $u_i$ and $v_i$, similarly to \cite{RYB19,OME13,STO00,LOW03,LOW06}. The strength of the diffusive coupling is given by $\sigma_u$ and $\sigma_v$ for the activator and the inhibitor variable, respectively. The coupling function of a given FHN oscillator $i$ is normalized by the number of oscillators coupled to it, i.e., its node degree $d_i=\sum_{j=1}^N A_{ij}$.  

\begin{figure}[!htp]
\includegraphics[width=\textwidth]{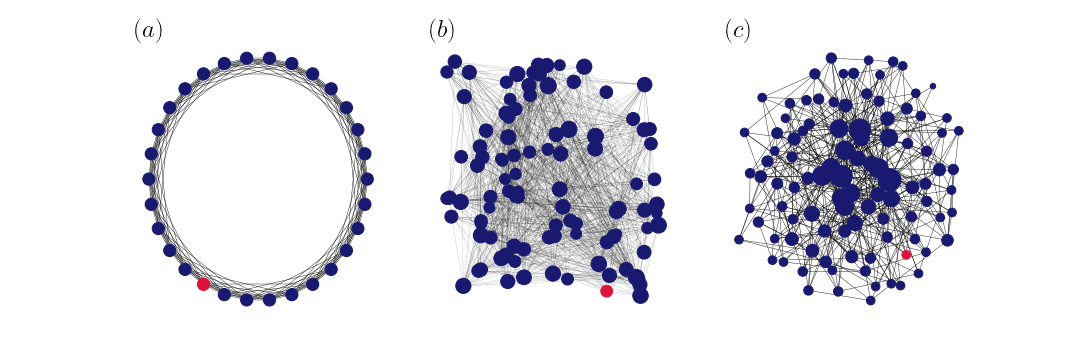}
\caption{Example of a solitary state in: (a) a nonlocally coupled ring graph (for better visibility we use $N=30$ and $R=5$, yielding $d_i=d=10$); (b) an Erd\H{o}s-R\'{e}nyi graph with $N=100$, $p=0.18$, yielding an average node degree of $\langle d_i \rangle \approx 18$; (c) Barab\'{a}si-Albert graph with $N=100$ and $m=5$ links added at each algorithm step, yielding $\langle d_i \rangle \approx 9.5$. The size of the nodes is proportional to the respective node degree.}
\label{fig:figure1}
\end{figure}

In our study, we consider three different network topologies, namely, a symmetric nonlocally coupled ring, a random, and a scale-free network. The topology of the symmetrically coupled ring is determined by a coupling radius $R$ common to all nodes in the network. The coupling radius $R$ defines the number of neighbors in each direction on a ring. This coupling scheme yields a uniform node degree $d_i=2R$ $\forall$ $i \in [1,N]$, see the corresponding graph in Fig. \ref{fig:figure1}(a). This symmetric case is used as a reference case for the comparison with the asymmetric structures. To obtain the random topology, we consider the well-known Erd\H{o}s-R\'{e}nyi (ER) algorithm \cite{BOL98}. Starting from $N=100$ unconnected nodes, we establish a connection between any given pair of nodes with a probability $p=0.18$. This procedure results in a normal distribution of node degree with average value of $\langle d_i \rangle = 18$, see the corresponding graph in Fig. \ref{fig:figure1}(b). Next, in order to obtain a scale-free network, we use the Barab\'{a}si-Albert (BA) algorithm which considers preferential attachment in the network growth process \cite{ALB02,BAR13}. Starting from a network with a star-like topology, each step of the BA algorithm consists of adding $m$ new links into such a network. The newly added links are distributed across the network as follows: a given node $i$ receives a new link in accordance with its current degree $d_i$ with the following probability $p_i = \frac{d_i}{\sum_i d_i}$. This procedure yields a network with a power-law distribution of node degree. In Fig. \ref{fig:figure1}(c), we show the resulting graph of the BA algorithm with $m=10$. 

\begin{figure}[!htp]
\includegraphics[width=\textwidth]{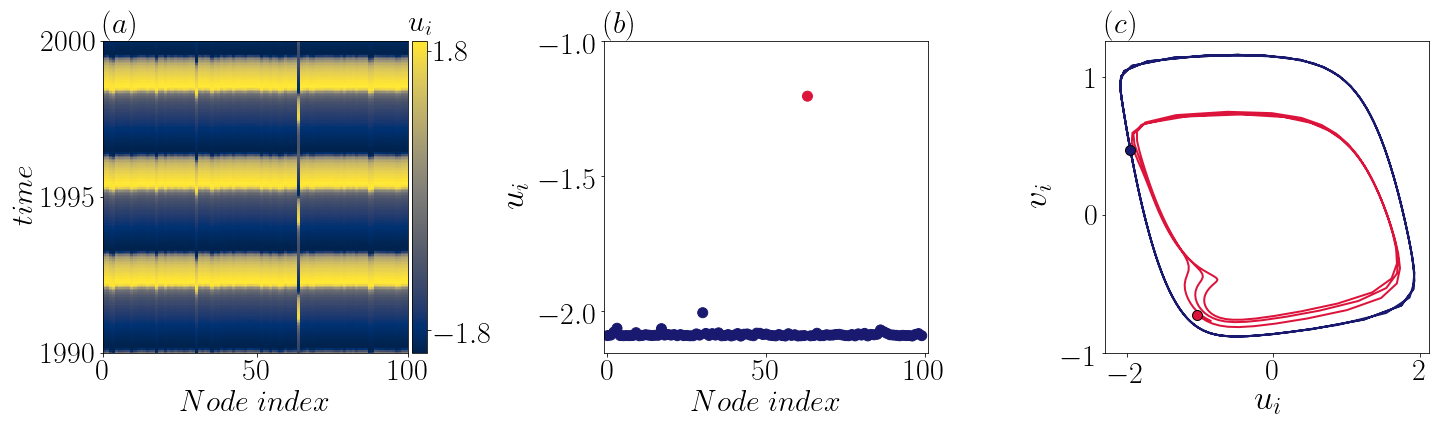}
\caption{Solitary state with a single solitary node in a scale-free network with $N=100$, $m=5$ (as shown in Fig. \ref{fig:figure1}c), $\sigma_u=0.12$, and $\sigma_v=0.15$: (a) Space-time plot for the activator variable $u_i$; (b) Snapshot of the activator variable $u_i$ taken at $t=2000$ (arb. units). Blue circles correspond to the synchronized cluster, while the red one shows the solitary node; (c) State-space projections for the solitary node (red) and the synchronized cluster (blue).}
\label{fig:figure2}
\end{figure}

We now investigate the onset of solitary states in the considered asymmetrically coupled networks. For that, we attribute random initial conditions (ICs) to the system in Eq. (\ref{equ:fhn_network}) with the adjacency matrix $A_{ij}$ first prescribing the scale-free topology shown in Fig. \ref{fig:figure1}(c). With this, in Fig. \ref{fig:figure2}(a), we show the time evolution of the network, illustrating the presence of one solitary node split off from the main synchronized cluster. This behavior can be further visualized in a snapshot of the variable $u_i$ at $t=2000$ (arb. units) on display in Fig. \ref{fig:figure2}(b). In general, the trajectories of solitary nodes approach an attractor coexisting with the one hosting the synchronized cluster. The attractor of a solitary node can be periodic or chaotic \cite{RYB19,SCH22}. For the case depicted in Figs. \ref{fig:figure2}(a) and \ref{fig:figure2}(b), the dynamics of the solitary node follows a period-$3$ limit cycle as shown in the state-space projection ($u_i$,$v_i$) depicted in Fig. \ref{fig:figure2}(c). This figure demonstrates the existence of solitary states in asymmetric networks.  

\begin{figure}[!htp]
\includegraphics[width=\textwidth]{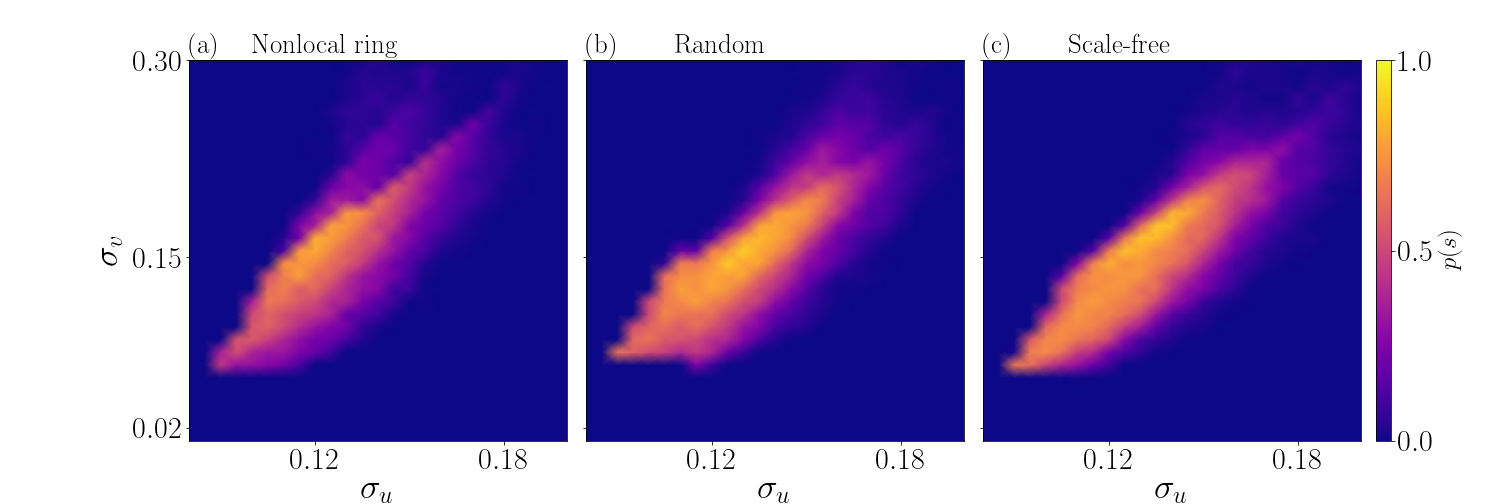}
\caption{Map of regimes in the ($\sigma_u, \sigma_v$)-parameter plane. Color-coded is the probability of obtaining a solitary state for the following topologies: (a) A nonlocally coupled ring network with $R=9$; (b) A random (Erd\H{o}s-R\'{e}nyi) network with average degree $p=0.18$; (c) A scale-free (Barab\'{a}si-Albert) network with $m=10$. The probability is estimated from $100$ network realizations with different ICs uniformly distributed over the intervals $u_i \in [-2.2,2.2]$, $v_i \in [-1.1,1.1]$. The network size is $N=100$ in all cases.}
\label{fig:figure3}
\end{figure}

In order to establish the coupling intensities for which solitary states arise in complex networks, we obtain the map of regimes in the ($\sigma_u, \sigma_v$)-parameter plane for all three topologies shown in Fig. \ref{fig:figure1}. For that, we consider an ensemble of $100$ network realizations with different ICs. For each simulation, we analyze a time interval of $\Delta t=2000$ (arb. units) for establishing the occurrence of solitary states. This approach provides statistical results in which the fraction
of simulations resulting in solitary states is characterized by the probability of their occurrence. With this, in the map of regimes shown in Figs. \ref{fig:figure3}(a)-(c) for the respective topologies, the color-code stands for the probability of obtaining a solitary state for a given parameter pair ($\sigma_u, \sigma_v$). We observe that the parameter regions with a high probability of finding solitary states occur for very similar values of the coupling intensities $\sigma_u$ and $\sigma_v$ for all three topologies. In addition, the shape of such parameter regions is similar across the topologies. Therefore, the analysis of the ($\sigma_u, \sigma_v$)-parameter plane does not capture the influence of topological asymmetries. To tackle this issue, we analyze in more detail the topological features of networks exhibiting solitary states in the subsequent sections.

To complement our analysis of the coupling intensities, we point out that in a recent work for a globally coupled system (symmetric), Sch\"ulen et al. have obtained similar patterns in the ($\sigma_u, \sigma_v$)-parameter plane \cite{SCH22}. For a class of such systems, the authors have derived the bifurcation curves delimitating the regions of high probability for the occurrence of solitary states \cite{SCH22}. The similarity of the patterns observed for the globally coupled system and the networks studied here suggests that the same bifurcation scenario gives rise to solitary states in the networks with complex topologies. This fact is also investigated subsequently in this work. 

\section{The role of the network connectivity and local topological features}

\begin{figure}[!htp]
    \centering
    \includegraphics[width=\textwidth]{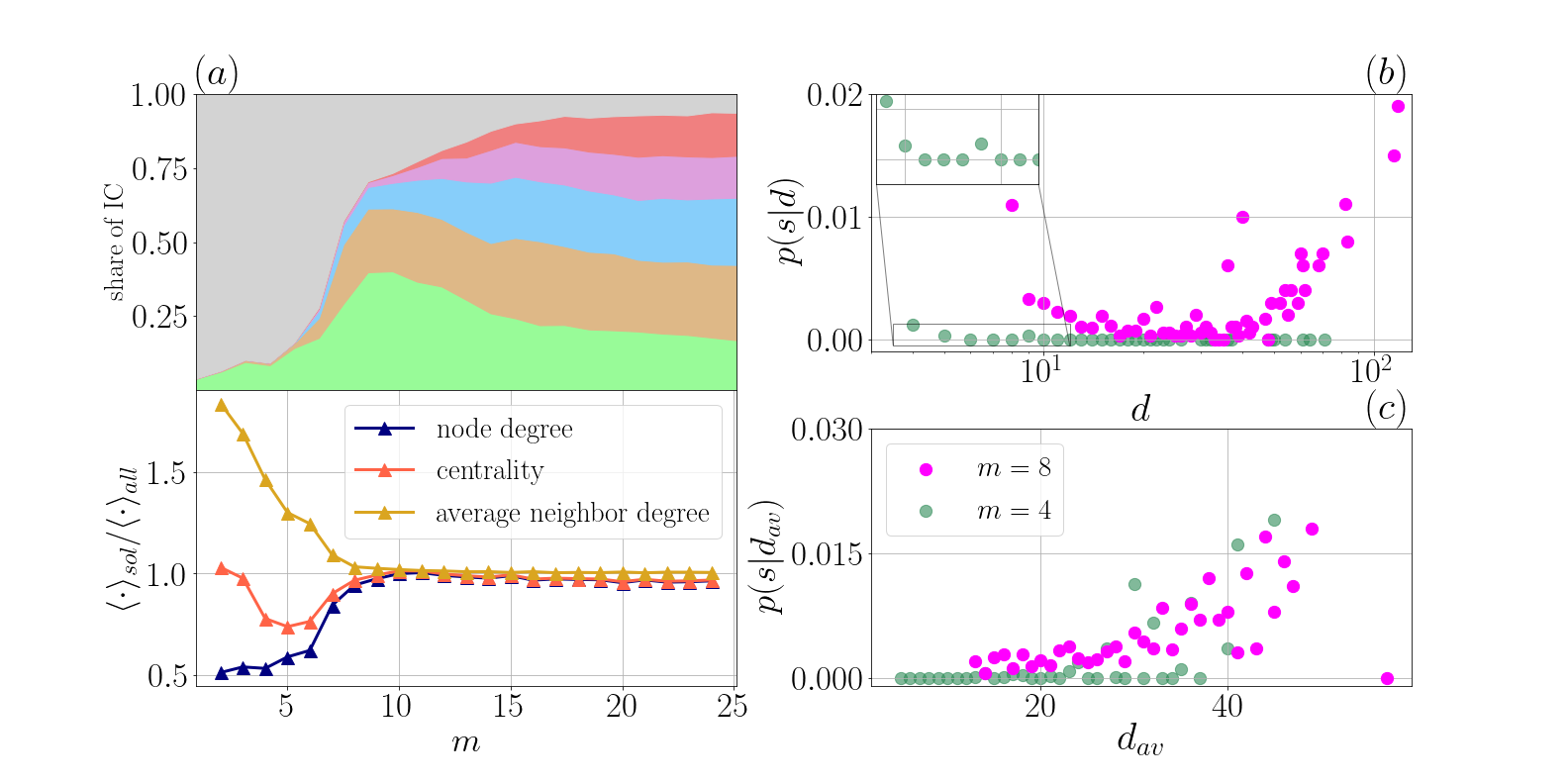}
    \caption{(a) Top panel: Share of network realizations yielding a solitary state with one (green), two (orange), three (blue), four (light purple), and five or more (red) solitary nodes for $m \in [1,25]$. The gray color corresponds to realizations with no solitary states. Bottom panel: Ratio $\langle . \rangle_{sol} / \langle . \rangle_{all}$ between average network measures, node degree (blue), centrality (red), and average neighbor degree (yellow). Parameters are: $\sigma_u=0.12$, $\sigma_v=0.12$, and $N=100$. (b) Conditional probability $p(s | d)$ to obtain a solitary state for a given node degree $d$ in a Barab\'{a}si-Albert network with $m=4$ (green) and $m=8$ (magenta). (c) Conditional probability $p(s | d_{av})$ to obtain a solitary state for a given average neighbor degree $d_{av}$ in a Barab\'{a}si-Albert network with $m=4$ (green) and $m=8$ (magenta). Parameters in (b) and (c): $\sigma_u=0.12$, $\sigma_v=0.15$, and $N=500$.}
    \label{fig:figure4}
\end{figure}

The results conveyed in Sec. 2 show that solitary states arise in networks with complex topologies for the same range of coupling strength as for simple regular networks. This raises a general question: What would be the topological requirements for the onset of the solitary states? In order to investigate the existence of possible constraints, we concentrate our efforts on networks with topologies represented by BA graphs as they constitute the example with asymmetric distribution of node degrees. As discussed in Sec. 2, the connectivity of BA graphs is specified by the number of links $m$ added at every step of the growth algorithm. With this, we investigate the onset of solitary states in such networks for different levels of connectivity by varying $m$. For that, we first adopt a statistical approach by considering $1000$ realizations of the system Eqs. (\ref{equ:fhn_network}) with different ICs chosen randomly as discussed in Fig. \ref{fig:figure3}. In the top panel of Fig. \ref{fig:figure4}(a), we show the share of ICs resulting in solitary states with different number of solitary nodes for varying $m$ as follows: one (green), two (orange), three (blue), four (orchid) and five or more (red). The absence of solitary nodes (regime of synchronization) is marked in gray. The colored region is the relative fraction that of all ICs that goes to this specific state. For example, if $m=10$, about $35 \%$ of the simulations go to a state with a single solitary oscillator, about $60 \%$ go to either a single or two solitary nodes, so the relative share of two solitaries is about $25\%$. Thus, they are given by the differences between two curves. In effect, we have a $100 \%$ stacked bar for each value of $m$. For low values of $m$ (low connectivity), we observe that solitary states occur for a small share of the ICs and they contain only one solitary node. By increasing $m$, the share of realizations corresponding to the completely synchronized network (gray) diminishes giving room to a larger variety of solitary states. For instance, $m \approx 8$, the solitary states are prevalent among all the realizations being observed with a different number of solitary nodes. Despite the fact that they are observed for all levels of network connectivity, the findings in Fig. \ref{fig:figure4}(a) indicate that high connectivity favors the formation of such states. 

Next, we refine our knowledge by analyzing how higher connectivity influences the formation of solitary states. Specifically, we investigate local node properties to find out if the success of high network connectivity can be captured by measures such as the node degree, average neighbor node degree, and eigenvector centrality. As the node degree $d_i$ has been already defined in Sec. 2, the average neighbor node degree of node $i$ is given by $d_{av,i}=\frac{\sum_{j=1}^{N} A_{ij} d_j}{d_i}$. This measure accounts for the average level of connectivity of adjacent nodes of $i$. The eigenvector centrality of a node $i$ is given by $c_i = \frac{1}{\lambda} \sum_{j=1}^{N} A_{ij} c_j$, where $\lambda$ is the largest eigenvalue calculated via $\mathbf{A}\mathbf{c}=\lambda \mathbf{c}$. This measure ranks the node $i$ following the connectivity of its neighbors. With this, we first obtain the average of these three measures over the entire network, denoted by $\langle . \rangle_{all}$. Next, we obtain the average only over the solitary nodes, which we denote by $\langle . \rangle_{sol}$. We calculate the ratio $\langle . \rangle_{sol} / \langle . \rangle_{all}$ as a function of the parameter $m$ in the bottom panel of Fig. \ref{fig:figure4}(a). Interestingly, for $m \lesssim 8$, the average over solitary nodes differs from the network average. In particular, we point out the fact that the average neighbor node degree is much higher for the solitary nodes, i.e., $\langle d_{av,i} \rangle_{sol}>\langle d_{av,i} \rangle_{all}$ (yellow curve in the bottom panel of Fig. \ref{fig:figure4}(a)). This measure will be justified later on by the bifurcation analysis (Sec. 4). In addition, by comparison with the top panel of Fig. \ref{fig:figure4}(a), we conclude that the asymmetries between the solitary nodes and the nodes in the synchronized cluster favor the appearance of solitary states with only one solitary node (green region in the top panel of Fig. \ref{fig:figure4}(a)). In contrast, at $m \approx 8$, the ratio $\langle . \rangle_{sol} / \langle . \rangle_{all}$ of all measures approaches one, indicating the establishment of homogeneity among the solitary and the synchronized nodes. This configuration holds for $m > 8$, giving rise to the regime dominated by solitary states with a large number of solitary nodes. 

To further analyze the dependencies observed in Fig. \ref{fig:figure4}(a), we now investigate the probability of a node being a solitary given one of its characteristics discussed above such as node degree, average neighbor node degree, eigenvector centrality. For that, we employ the concept of conditional probability, given by
\begin{equation}
 p(s | x) = \frac{p(s \cap x)}{p(x)},
\end{equation}
where $p(s | x)$ stands for the probability of a node being a solitary $s$ given a topological property $x$, e.g., the node degree $d$ or the average neighbor node degree $d_{av}$. To calculate this conditional probability, we consider $1000$ realizations with different ICs of BA networks possessing $N=500$ nodes (for better statistical quality). With this, in the top panel of Fig. \ref{fig:figure4}(b), we obtain the conditional probability of a node being a solitary given its node degree, i.e., $p(s | d)$. For $m=4$ (green circles), we observe that only nodes with low $d<10$ degree have a nonzero probability of being solitary (see the inset in the top panel of Fig. \ref{fig:figure4}(b)). The probability of occurrence of solitary nodes with a degree above this threshold is zero. This suggests that the basins of attraction of the solitary states are rather small and does not necessarily mean that solitary states are impossible for these node degrees. On the other hand, for $m=8$ (pink circles), we find that both nodes at the lower and the higher end of the degree distribution have a high probability of becoming a solitary node. This inconsistency between the cases $m=4$ and $m=8$ suggests that the probability of a node being solitary is not only determined by its respective degree. Therefore, we now look into the conditional probability considering the average neighbor degree as the underlying condition $p(s | d_{av})$. In Fig. \ref{fig:figure4}(c) for both cases ($m=4$ and $m=8$), we observe that a high average neighbor degree corresponds to a higher probability of the respective node being solitary. This observation is in line with the findings in the bottom panel of Fig. \ref{fig:figure4}(a). Note that for $m \lesssim 8$ the average neighbor degree is the only measure that is on average higher for the solitary nodes with respect to mean of the network, i.e. $\langle d_{av,i} \rangle_{sol}>\langle d_{av,i} \rangle_{all}$. As $m$ increases, the high connectivity in the network provides a higher average neighbor degree to more nodes, creating the abundance of solitary states observed in the top panel of Fig. \ref{fig:figure4}(a).

\section{Bifurcation analysis}

\begin{figure}[!htp]
    \centering
    \includegraphics[width=\textwidth]{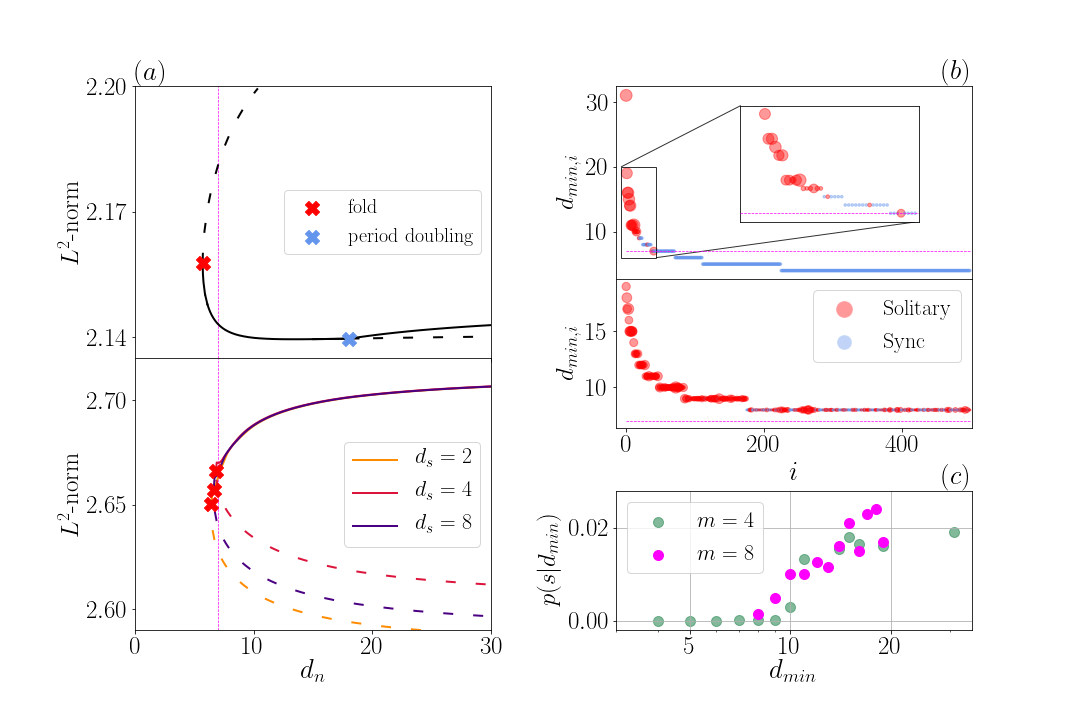}
    \caption{(a) Bifurcation diagram of the reduced systems varying $d_n$ as a bifurcation parameter. Red crosses indicate saddle-node bifurcations. The vertical dashed line in magenta indicates the threshold value $d_c=7$. Top panel: Reduced system in Eqs. (\ref{Equ:degreduced}). Bottom panel: Reduced system in Eqs. (\ref{Equ:threeReduced}) for $d_s=2$ (yellow), $d_s=4$ (red) and $d_s=8$ (purple). (b) The minimal neighbor degree of all nodes in Barab\'{a}si-Albert network with $m=4$ (top panel) and $m=8$ (bottom). Red circles indicate nodes that are observed as solitaries. Blue circles are the nodes that are always in sync. The size of the red circles is proportional to how often the corresponding node is observed as a solitary. (c) Conditional probability $p(s|d_{min})$ to observe a solitary for a given minimal neighbor node degree for Barab\'{a}si-Albert networks with $m=4$ (green) and $m=8$ (pink). Parameters: $\sigma_u=0.12$, $\sigma_v=0.15$, and $N=500$.}
    \label{fig:figure5}
\end{figure}

The statistical analysis performed in Sec. 3 has shown that the average neighbor node degree plays an important role in the formation of solitary states. Now we deepen this knowledge by investigating the bifurcation scenario giving rise to the solitary states in the considered complex network. To this end, we make use of the fact that the entire synchronized cluster influences the solitary node as a single input. That is, if node $s$ is a solitary and it is connected to $d_s$ oscillators that are all in sync with each other, the resulting coupling term in $s$ is given $C_s=\sigma_u (u_{b} - u_s) + \sigma_v (v_b - v_s)$, where $(u_b, v_b)$ are the state variables of oscillators in the synchronized cluster. In turn, an oscillator $n$, with node degree $d_n$, belonging to the synchronized cluster, and directly coupled to the solitary node, has nonzero coupling input only from the solitary node. The resulting coupling term in $n$ reads $C_n=1/d_n \cdot [\sigma_u (u_s - u_n) + \sigma_v (v_s - v_n)]$. With this, one can immediately infer that for large values of $d_n$, the coupling term for the node $n$ vanishes and it decouples from the solitary node $s$. Therefore, in order to ensure the coupling between the nodes $s$ and $n$, we assume that $n$ has the minimum node degree $d_n$ among all neighbors of $s$. In addition, we also assume that the rest of the network stays synchronized at all times. With these assumptions, we can reduce the dynamics of the network to the following equation of two interacting nodes \cite{GOL2003}: 
\begin{align}
\label{Equ:degreduced}
    \varepsilon \frac{du_s}{dt} &= f(u_s, v_s) + \sigma_u (u_n - u_s) + \sigma_v (v_n - v_s) \nonumber, \\
    \frac{dv_{s}}{dt}&=g(u_s, v_s) \nonumber, \\
    \varepsilon \frac{du_n}{dt} &= f(u_n, v_n) + \frac{1}{d_n}[\sigma_u (u_s - u_n) + \sigma_v (v_s-v_n)], \nonumber \\
    \frac{dv_{n}}{dt}&=g(u_n, v_n),
    \end{align}
where the functions $f(u,v)$ and $g(u,v)$ are given in Eqs. (\ref{equ:functions}). Following this dimension reduction, we are now able to study the stability of solutions of the system Eqs. (\ref{Equ:degreduced}). Considering the node degree $d_n$ as a continuous parameter, we employ numerical continuation analysis to follow a solitary solution as $d_n$ is varied. For this task, we use the software \textbf{auto-07p} \cite{DOE07}. The resulting bifurcation diagram is shown in the top panel of Fig. \ref{fig:figure5}(a). Even though, in reality the minimum node degree $d_n$ is an integer number, the bifurcation analysis in Fig. \ref{fig:figure5}(a) treating it as continuous provides deep insight into the stability profile of the solitary node $s$. Note that the limit-cycle hosting the trajectory of the solitary node appears in a fold bifurcation at $d_n \approx 5.701$, see the red cross in the top panel of Fig. \ref{fig:figure5}(a). This analysis indicates that the neighbor nodes of a solitary node must have a minimum node degree of $d_n \geq 6$. Therefore, we establish this value of $d_n$ as a threshold value $d_c$ for the occurrence of solitary states in the network. In addition, it is also visible in the top panel of Fig. \ref{fig:figure5}(a) that the increase of the parameter $d_n$ leads to a period-doubling bifurcation transforming the dynamics of the solitary node, see a blue cross. This observation suggests that the larger the neighbor node degree, the higher is the period of the limit-cycle hosting the solitary node.

The results shown in the top panel of Fig. \ref{fig:figure5}(a) are obtained by assuming the dynamics of two coupled oscillators as a descriptor of two interacting subsets of the network, i.e., the synchronized cluster and the solitary node. This approach can be improved by splitting the synchronized cluster into the dynamics of a mean-field component and the dynamics of the node directly coupled to the solitary node. With this, the dynamics of the solitary node $s$ and the directly coupled neighbor $n$ described in Eqs. (\ref{Equ:degreduced}) receive an extra dynamical input corresponding to the synchronized mean-field. The dynamics of the mean-field is described by an extra oscillator with state variables $(u_b, v_b)$, which are not affected by the nodes $n$ and $s$. The resulting equations for this version of the reduced system are given by: 
\begin{eqnarray}
    \varepsilon \frac{du_b}{dt} &=& f(u_b, v_b) , \qquad \frac{dv_{b}}{dt} = g(u_b, v_b) \nonumber, \\
    \varepsilon \frac{du_n}{dt} &=& f(u_n, v_n) + (1-\frac{1}{d_n}) [\sigma_u (u_b - u_n) + \sigma_v (v_b - v_n)] + \frac{1}{d_n} [\sigma_u (u_s - u_n) + \sigma_v (v_s - v_n)] \nonumber, \\
     \frac{dv_{n}}{dt} &=& g(u_n, v_n), \nonumber \\
    \varepsilon \frac{du_s}{dt} &=& f(u_s, v_s) + (1-\frac{1}{d_s}) [\sigma_u (u_b - u_s) + \sigma_v (v_b - v_s)] + \frac{1}{d_s} [\sigma_u (u_n - u_s) + \sigma_v (v_n - v_s)], \nonumber \\
    \frac{dv_{s}}{dt} &=& g(u_s, v_s),
    \label{Equ:threeReduced}
\end{eqnarray}
with $f(u,v)$ and $g(u,v)$ as in Eqs. (\ref{equ:functions}). As before, the bifurcation analysis is performed by varying the minimum degree $d_n$ of the neighbor node, see the bottom panel of Fig. \ref{fig:figure5}(a). We repeat the analysis for three different values of the parameter attributed to the degree of the solitary node, namely $d_s \in [2,4,8]$. Surprisingly, the choice of $d_s$ does not influence much the threshold for the existence of solitary states. In all three cases, we find the fold bifurcation located between $6<d_n<7$, slightly higher compared to the case of the two-node reduction. This implies that the lower threshold is $d_n \geq 7$. We can therefore confirm the claim that the crucial parameter for the formation of solitary states is the minimal neighbor node degree $d_n$. Neither the node degree of the solitary node nor the \emph{average} neighbor node degree by itself have a causal role in the formation.

Finally, we statistically verify this result in the full system, i.e., via an ensemble of realizations of the complex network. For that, we consider realizations of BA networks with two different values of the parameter $m$ to compute the quantity $d_{min,i}$, the minimum neighbor node degree of each network node $i$. For $m=4$, in the top panel of Fig. \ref{fig:figure5}$(b)$, we obtain $d_{min,i}$ for all nodes $i$. In this figure, among the different realizations, the nodes that are solitary at least once are colored in red, while the other nodes are marked in blue. In addition, the size of the red circles is proportional to how often the corresponding node is observed as a solitary. We observe that the nodes with a higher value of $d_{min,i}$ are more often solitaries, indicating that indeed the higher value of the minimum neighbor node degree yields a higher probability of being solitary. Moreover, for the $m=4$, the BA network has low connectivity implying that most nodes have a $d_{min}$ below the threshold $d_c=7$ unraveled by the bifurcation analysis in Fig. \ref{fig:figure5}(a). The dashed horizontal line in Fig. \ref{fig:figure5}(b) marks this threshold and indeed all nodes below this threshold do not become solitary (blue circles). Conversely, in a BA network with $m=8$, all nodes have a degree larger than the critical threshold $d_c$. In such a network, all nodes meet the criteria we established via bifurcation analysis. The network simulations shown in the bottom panel of Fig. \ref{fig:figure5}(b) confirm this hypothesis. Note that $d_{min}$ for all nodes falls above the dashed horizontal line marking $d_c$ and most of the nodes can be solitaries (red circles). Also, in this case, higher $d_{min}$ corresponds to a higher probability of being solitary. Finally, in Fig. \ref{fig:figure5}(c), we complement our analysis by estimating the conditional probability $p(s|d_{min})$ of a node $i$ with a given $d_{min,i}$ being solitary. For $m=4$ (green circles), the observed probability is nonzero only for the nodes having $d_{min,i}>d_c$. In addition, the probability increases for nodes with higher values of $d_{min,i}$. For $m=8$ (pink circles), as all nodes already fall into the criteria for solitary ($d_{min,i}>d_c$ $\forall$ $i \in [1,500]$), there is no node with zero probability of being a solitary. Again, the probability is higher for nodes with higher $d_{min,i}$. Both cases confirm the existence of the threshold $d_c$ for the minimum node degree of adjacent nodes provided by the bifurcation analysis.

\section{Discussion}

In summary, by approaching the onset of solitary states in complex networks, we have demonstrated that the connectivity of nodes adjacent to a solitary one is an essential topological feature for the appearance of solitary states. More specifically, we found a threshold for the minimum value of the degree of nodes neighboring the solitary ones in the network. This finding is statistically demonstrated by estimating the conditional probability of its occurrence in ensembles of realizations of scale-free networks. Furthermore, a dimensionality reduction of the network dynamics made possible a bifurcation analysis confirming the existence of the minimum threshold.    

Due to the symmetry-broken character of solitary states, their occurrence is undesirable in many contexts. Therefore, the knowledge of topological constraints for the onset of these states sheds light on a perspective of their control, i.e., suppression or initiation. Finally, we emphasize that the results reported here can be extended to other classes of networked systems since there are no particular dynamical, or topological, restrictions on them.  
\vspace{1cm}

Code and data are accessible upon request. \\

We thank Matthias Wolfrum and Alexander Gerdes for fruitful discussions. E.S.M acknowledges the support by the Deutsche Forschungsgemeinschaft (DFG) via the project number FE-359/22-1. This work was supported by the Deutsche Forschungsgemeinschaft (DFG, German Research Foundation) - Projektnummer - 163436311 - SFB 910.

\bibliographystyle{unsrt}
\bibliography{refs}

\end{document}